\documentstyle[12pt]{article}
\begin{document}
\input{psfig}

\textheight 19cm

\newcommand{\ap}{\alpha '}
\newcommand{\del}{\Delta}

\begin{titlepage}
\rightline{hep-th/9903038}
\rightline{TAUP-2564-99}
\vskip 1cm
\centerline{{\Large \bf  Semi Localized Brane Intersections in SUGRA}}

\vskip 1cm
\centerline{
A. Loewy
\footnote{e-mail: loewy2@post.tau.ac.il}}

\vskip 1cm
\begin{center}
\em School of Physics and Astronomy
\\Beverly and Raymond Sackler Faculty of Exact Sciences
\\Tel Aviv University, Ramat Aviv, 69978, Israel

\end{center}
\vskip 1cm

\begin{abstract}
We construct SUGRA solutions of brane configurations of intersecting
Dp and NS branes. These solutions are semi-localized in the sense that one of
the intersecting branes is smeared along the world volume of the
other, while the second is localized. We examine the gauge theory
that lives on the localized brane, and the various descriptions
possible via T and S dualities, and M-theory.
\end{abstract}
\end{titlepage}


\section{Introduction}
\label{sec:1}
The SUGRA description of intersecting p-branes was considered by many
authors \cite{gaun,tow,tow1,tsey,tsey1,tsey2,englert,boonstra}. In all these
works intersecting branes were smeared in
their relative transverse directions. The SUGRA solution was
characterized by two harmonic functions of the overall transverse
coordinates. Intersection rules have been derived in more than one way
to ensure stability, and some unbroken super-symmetry \cite{englert,tsey1}. The
metric, dilaton, RR forms and the NS-NS 2-form are given by a
superposition rule \cite{tsey2}. We
specialize to the case where a Dp-brane intersects an NS brane over
p-1 dimensions. 
\begin{eqnarray}
  \label{hsr}
  ds^{2} & = & H_{p}^{-1/2} dx_{0..p-1}^{2} + H_{p}^{1/2} dx_{p..5}^{2} + H_{p}^{-1/2} H_{5} dx_{6}^{2} + H_{p}^{1/2} H_{5}
  dx_{789}^{2} \nonumber \\
e^{2 (\phi-\phi_{\infty})} & = & H_{5} H_{p}^{(3-p)/2} \nonumber \\
C_{01..p} & = & H_{p}^{-1} \ \ \ \ \  dB_{2}=*dH_{5}
\end{eqnarray}
where $H_{p}$ and $H_{5}$ are the Dp and NS brane harmonic
functions for $p \le 4$, $x_{0..p-1}$ and $x_{p..5}$ are the coordinates
shared by both branes and the coordinates that belong to the NS brane
but not to the Dp-brane respectively. The RR form flux is quantized,
and corresponds to the number of Dp-branes. We 
generalize this construction and allow either $H_{p}$ or $H_{5}$ to depend on relative
transverse coordinates as well. The function describing the localized
brane  should obey
a curved space Laplace equation \cite{tsey2,sanny,yang}, which in the Einstein frame is
\begin{equation}
  \label{csle}
  g^{\mu \nu} \partial_{\mu} \partial_{\nu} H_{i} = 0
\end{equation}
Note that the functions will not depend on coordinates in the
directions shared by both branes (i.e. 0..p-1).
The resulting solution describes a semi localized intersection of two
branes. Finding such a solution involves solving the nonlinear
differential equation, (\ref{csle}). Solutions of brane intersections in M-theory also obey a similar superposition rule.

SUGRA solutions of various brane constructions are of interest due to
the AdS/CFT duality. Intersecting brane configurations were found to
be powerful tools in analyzing supersymmetric gauge dynamics in
various dimensions \cite{witten,gk}. In \cite{malda,cobi} various brane solutions with 16
supersymmetries and their field theory duals where discussed. By
finding SUGRA solutions for intersecting brane configurations one can
make statements about field theories with less supersymmetries. The
near horizon behavior of such solutions may give us a SUGRA dual of the corresponding
field theory. However, if the solution is that of a smeared
intersection then the field theory data (gauge couplings, energy scale) is
not apparent.  

In this paper we shall
find semi-localized solutions, that is solutions in which one of
the harmonic functions depends only on the overall transverse coordinates
and the other depends on relative transverse coordinates as well. We
shall comment on the field theory duals that these solutions describe.
The general form of such solutions was also considered by
\cite{tatar,bert,yang}. Some
of the solutions that will be presented here have also been obtained by
dimensional reduction from 11 to 10 dimensions in the presence of
singularities \cite{sanny,hashimoto,oscar}.

While finishing this work two related papers by D. Youm hep-th/9902208, and A. Fayyazuddin, D. Smith hep-th/9902210, that
contain overlapping results appeared in the electronic archive.

\section{Semi-localized NS branes}
\label{sec:2}
We start with solutions of Dp $\bot$ NS intersections where the NS
brane is smeared along the 6 direction, and has its world volume along
(12345). We shall think of this configuration as the continuum limit of a periodic array of NS branes
along 6, making $H_{5}$ independent of $x_{6}$, see fig.(1). In order that the
total number of NS branes be finite, we shall compactify $x_{6}$, and
probe this configuration only at distances larger than the
compactification radius. We
denote by $v$ the coordinates that belong to the NS but not to the
Dp-brane, which is extended in (12..p-1,6).
\begin{figure}[htbp]
  \begin{center}
    \leavevmode
    \psfig{figure=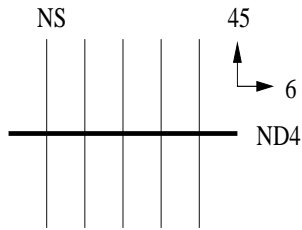,width=4cm,height=3cm,clip=}
    \label{fig:1}
    \caption{A smeared NS-brane as the continuum limit of a periodic
      array of NS-branes. The D4-brane is stretched along the compact
      coordinate 6. Other Dp-brane configurations are similar.}
  \end{center}
\end{figure}
\subsection{NS $\bot$ D4}
We start with an intersection of an NS and D4. The harmonic function
describing $N_{5}$ smeared NS
branes satisfies $\del_{789} H_{5}=0$. 
This is trivially solved by $1+Q_{5}/r$, where
$r^{2}=x_{7}^{2}+x_{8}^{2}+x_{9}^{2}$ (we consider only the spherical
symmetric solution for reasons explained below). We shall work in the
near horizon limit $H_{5}=Q_{5}/r$. Thus the curved space Laplace
equation for the harmonic function describing the $N_{4}$ D4-branes, $H_{4}$ is
\begin{equation}
  \label{laph4}
 \Big[ \del_{45} + H_{5}^{-1} \del_{789} \Big] H_{4}(v,r) \ = \ 0
\end{equation}
By defining $u=2\sqrt{Q_{5}r}$, the above equation becomes a sum of
two radial Laplace operators in flat space
\begin{equation}
  \label{laph4f}
  \Big[ v^{-1}\partial_{v}(v\partial_{v}) +
  u^{-2}\partial_{u}(u^{2}\partial_{u}) \Big] H_{4}(v,u) \ = \ 0
\end{equation}
Eq.(\ref{laph4f}) can be further simplified
by defining $w^{2}=v^{2}+u^{2}$.
\begin{equation}
  \label{nsd4w}
  \Big[ w^{-5}\partial_{w}(w^{5}\partial_{w}) + w^{-2}
  \Lambda(\lambda) \Big] H_{4}(w,\lambda) \ = \ 0
\end{equation}
where $\tan \lambda=u/v$, and $\Lambda(\lambda)$ is the angular part
of the Laplace operator. A solution with no $\lambda$ dependence is
$H_{4}=1+Q_{4}/w^{4}$ which in the near horizon limit, $Q_{4} \gg w^{4}$, gives according
  to eq.(\ref{hsr}) the following SUGRA solution 
\begin{eqnarray}
  \label{nsd4met}
  ds^{2} & = &  \frac{w^{2}}{\sqrt{Q_{4}}} dx^{2}_{0123} +
  \frac{\sqrt{Q_{4}}}{w^{2}} dw^{2} +\sqrt{Q_{4}} \Big[d \lambda^{2} + \cos^{2}\lambda d
  \beta^{2} + \frac{1}{4} \sin^{2}\lambda d \Omega_{2}^{2} \Big] + \frac{4
    Q_{5}^{2}}{\sqrt{Q_{4}} \sin^{2} \lambda} d x_{6}^{2}  
  \nonumber \\
  e^{2(\phi-\phi_{\infty})} & = & H_{5} H_{4}^{-1/2} = \frac{4Q_{5}^{2}}{\sqrt{Q_{4}}
    \sin^{2} \lambda} \nonumber \\
R & \sim & \frac{1}{\sqrt{Q_{4}}}   
 \end{eqnarray}
where $R$ is the curvature of the AdS part of the metric.
 The field theory
described by this configuration should be an $N=2$, four dimensional 
$\Pi_{i=1}^{N_{5}} SU(N_{4})$ supersymmetric gauge theory with $N_{5}$ hyper-multiplets in the
bi-fundamental representation, $(N_{i},\bar{N}_{i+1})$, where
$i=N_{5}+1$ is identified with $i=1$, since $x_{6}$ is compact. There
is also an extra $U(1)$ factor, since this configuration is an
elliptic model \cite{witten1}. This field
theory is conformal ($N_{F}=2N_{C}$ for each of the gauge groups in the
product), as is obvious from the
SUGRA solution, which contains an AdS part.
We consider solutions that depend only on $r,v$ because we wish
to preserve the $SO(3)_{r} \times SO(2)_{v}$ isometry, which according
to the AdS/CFT duality should be identified with the R-symmetry of the
gauge theory.

We now want to write the SUGRA solution in terms of field theory
variables, namely the energy scale and $g_{YM}$.
\begin{equation}
  \label{q4q5}
  Q_{4} \sim g_{YM}^{2} N_{4} \ap^{2} \ \ \ \ \ Q_{5} \sim N_{5}
  \ap^{1/2} \ \ \ \ \ e^{2 \phi_{\infty}}=g_{s}^{2} \sim g_{YM}^{4}
\end{equation}
Since the field theory is 3+1 dimensional, $g_{YM}$ is
dimensionless. In order for the metric to have an $\ap$
factor in front, one should substitute $x_{6}^{2}=\ap \Phi^{2}$. We
also note that there seem to be two energy scales in the solution, $R=r/\ap$ and $V=v/\ap$. However, because $v$ and $r$ do not
enter symmetrically in the definition of $w$, we cannot at the same time
keep $R$ and $V$ fixed when taking $\ap \rightarrow 0 $. We will work
only with the combination of the two, $W=w/\ap$ and $\sin
\lambda$. Note, however, that $w=\sqrt{v^{2}+4Q_{5}r}$ does not have the
interpretation of the length of a stretched open string.

The 10 dimensional geometric description is only valid
when the effective  string coupling $e^{\phi}$, and the curvature of the
AdS part in string units are small. Thus as in
\cite{cobi}, we can deduce the energies at which we can use this dual
description. In this
case the dilaton and curvature are only functions of $\sin \lambda$,
which means that the geometric description is valid for: $e^{\phi} \ll
1$ and $\ap R \ll 1$.
\begin{equation}
  \label{nsd4con}
  e^{2 \phi} \sim \frac{g_{YM}^{3}N_{5}^{2}}{ \sqrt{N_{4}} \sin^{2}
    \lambda} \ \ \ \ \ \ap R \sim \frac{1}{g_{YM} \sqrt{N_{4}}}
\end{equation}
When $\sin \lambda \rightarrow 0$ we have two options. One is to
elevate this solution to M-theory. We get a configuration of two M5 branes, one of which is
smeared over the (6,10) directions, and localized in 789, and the other localized in
all its transverse directions (45,789). This solution is described
by the same harmonic functions as before, since the harmonic
superposition rule (M5 $\bot$ M5) in 11 dimensions gives the same
Laplace equation as the 10 dimensional one.
\begin{eqnarray}
  \label{nsd4m}
  ds^{2} & = & H_{4}^{-1/3}H_{5}^{-1/3} dx_{0123}^{2} +
  H_{4}^{2/3}H_{5}^{-1/3}dx_{45}^{2}+ H_{4}^{-1/3}H_{5}^{2/3}
  dx_{6,10}^{2} +H_{4}^{2/3}H_{5}^{2/3} dx_{789}^{2} \nonumber \\
l_{p}^{2} R & \sim & \Big[ \frac{N_{5}}{N_{4}\sin
  \lambda} \Big]^{2/3}
\end{eqnarray}
This description does not give anything new, since the curvature still
diverges when $\sin \lambda \rightarrow 0$.

We can also go to the T-dual configuration. By making a T duality in
the 6 direction we get $N_{4}$ D3 branes smeared in the 6 direction and
localized on a singular compact space, $X$. The new metric and dilaton are derived
from the original ones using T-duality rules \cite{berg}.
\begin{eqnarray}
  \label{nsd4td}
  ds^{2} & = & \frac{w^{2}}{\sqrt{Q_{4}}} dx_{0123}^{2} +
  \frac{\sqrt{Q_{4}}}{w^{2}} dw^{2} + \sqrt{Q_{4}} ds_{X}^{2} \nonumber \\
e^{2 \phi} & = & 1 
\end{eqnarray}
where $ds_{X}^{2}$ turns out to be a $Z_{N_{5}}$ orbifold of $S^{5}$,
where the orbifold acts on the $S^{3}$ (6789) part of eq.(\ref{nsd4met}). The connection
between the NS brane and $A_{N-1}$ singularities, which are described
by this metric, and the field
content of this theory were discussed in \cite{kachru,vafa,dasgupta,uranga}. In the T-dual picture
it is possible to generalize the solution to a near extremal brane in
the usual way by $g_{00} \rightarrow g_{00} f(w)$ and $g_{ww}
\rightarrow g_{ww} f^{-1}(w)$. One can now T-dualize back to the original
configuration and obtain the solution for the near extremal brane
intersection. This configuration is not expected to be stable. One
should note that the radial coordinate $w$ does not have the
interpretation of distance from the D3 in transverse space as in
Maldacena's original solution \cite{malda}, since $w^{2}=v^{2}+4Q_{5}r$.

\subsection{NS $\bot$ D3}
The next configuration we shall consider is an intersection of $N_{3}$
localized D3 branes with $N_{5}$ NS branes smeared over the 6
direction. The SUGRA solution can be derived in much the same way as in the
previous case. We know the near horizon behavior of $H_{5}=Q_{5}/r$
and thus the equation for $H_{3}$ is
\begin{equation}
  \label{nsd3la}
  \Big[ \del_{345} + H_{5}^{-1} \del_{789} \Big] H_{3}(v,r) \ = \ 0
\end{equation}
Again we substitute $u=2 \sqrt{Q_{5}r}$, look for solutions that
depend only on $u,v$ in order to keep the $SO(3)_{r} \times SO(3)_{v}$
isometry that will correspond to an R-symmetry, and write eq.(\ref{nsd3la}) as a 7 dimensional flat space Laplace operator.
\begin{equation}
  \label{nsd3f}
\Big[ w^{-6} \partial_{w}(w^{6} \partial_{w}) + w^{-2} \Lambda
(\lambda) \Big] H_{3} (w,\lambda) \ = \ 0  
\end{equation}
The solution that has no $\lambda$ dependence is $H_{3}=1+Q_{3}/w^{5/2}$ and
we shall work in the near horizon limit. The SUGRA solution follows, as
before, from eq.(\ref{hsr}) 
\begin{eqnarray}
  \label{nsd3met}
  ds^{2} & = & \frac{w^{5/2}}{\sqrt{Q_{3}}} dx_{012}^{2} +
    \frac{\sqrt{Q_{3}}}{w^{5/2}} d w^{2} + \frac{\sqrt{Q_{3}}}{w^{1/2}}
      \Big[ d \lambda^{2} + \cos^{2} \lambda d \Omega_{2}^{2} +
      \frac{1}{4} \sin^{2} \lambda d \Omega_{2}^{2} \Big] +
      \frac{4Q_{5}^{2}w^{1/2}}{\sqrt{Q_{3}} \sin^{2} \lambda} dx_{6}^{2} \nonumber \\
e^{2(\phi-\phi_{\infty})} & = & H_{5} = \frac{4Q_{5}^{2}}{w^{2} \sin^{2} \lambda}
\nonumber \\
R & \sim & \frac{\sqrt{w}}{\sqrt{Q_{3}}}
\end{eqnarray}
where $R$ is the curvature of the $(012w)$ part of the metric.
Note that, unlike the NS $\bot$ D4 solution, this metric does not
describe a product space with an AdS part, thus the corresponding
field theory will not be conformal.

 The field theory described by this configuration should be a
$\Pi_{i=1}^{N_{5}} SU(N_{3})$ gauge theory in d=2+1 dimensions with 8
super-symmetries and $N_{5}$ hyper-multiplets in the $(N_{i},
\bar{N}_{i+1})$ representations.  As in the previous case, there is an
extra $U(1)$ factor. 
The field theory description is valid where $g_{eff}^{2}=g_{YM}^{2}
N_{3} W^{-1}$ is small: $W \gg g_{YM}^{2} N_{3}$.

 The geometric dual is valid when the effective string coupling, as well as, the  curvature in string units, are small. To see that we first
write the solution in terms of the field theory variables, $W=w/\ap$ and $N_{3},N_{5}$.
\begin{equation}
  \label{q3q5}
  Q_{3} \sim g_{YM}^{2} N_{3} \ap^{3} \ \ \ \ \ Q_{5} \sim N_{5}
  \ap^{1/2} \ \ \ \ \ e^{2 \phi_{\infty}}=g_{s}^{2} \sim g_{YM}^{4}  \ap
\end{equation}
\begin{equation}
  \label{nsd3st}
  \frac{N_{5} g_{YM}^{2}}{\sin \lambda} \ < \ W \ < \
  g_{YM}^{2} N_{3} 
\end{equation}
When the string coupling becomes large we can go to the S-dual
description. The resulting SUGRA solution is just $H_{5}^{-1/2}$ times
the original, and the dilaton is just $- \phi$. 
\begin{eqnarray}
  \label{nsd3sd}
  ds^{2}_{dual} & = & H_{5}^{-1/2} ds^{2} \nonumber \\
  e^{2\phi} & = & H_{5}^{-1} \sim \frac{W^{2} \sin^{2}
    \lambda}{N_{5}^{2} g_{YM}^{4}} \nonumber \\
\ap R & \sim & \frac{N_{5} g_{YM}}{\sqrt{N_{3}}W^{1/2}}  
\end{eqnarray}
We can also T-dualize along the 6 direction thus getting $N_{3}$ D2 branes
smeared over 6, and localized on a singularity. The resulting SUGRA
solution is
\begin{eqnarray}
  \label{nsd3td}
  ds^{2} & = & \frac{w^{5/2}}{\sqrt{Q_{3}}} dx_{012}^{2} +
  \frac{\sqrt{Q_{3}}}{w^{5/2}} dw^{2} + \frac{\sqrt{Q_{3}}}{w^{1/2}}
  ds_{X}^{2}
 \nonumber \\
e^{2 \phi} & \sim & \frac{g_{YM}^{5} \sqrt{N_{3}}}{W^{5/2}}  
\end{eqnarray}
where now $ds_{X}^{2}$ is a $Z_{N_{5}}$ orbifold of $S^{6}$. As in the
D4 case, the $Z_{N_{5}}$ orbifold acts only on the $S^{3}$ (6789) part of
the transverse space. 
This solutions resembles the solution found in \cite{cobi} for D2 branes in
flat space, where the original $d \Omega_{6}$ is now orbifolded, and
$w$ is not the naive distance from the D2 brane. Again
we can generalize to a near extremal solution and T-dualize back to
get the near extremal solution of the intersection.

\subsection{NS $\bot$ D2}
The next configuration we shall discuss is the intersection of $N_{2}$
localized D2 branes with $N_{5}$ NS branes. $H_{5}$ is the same as in
the two previous cases, thus the equation for $H_{2}$ is
\begin{equation}
  \label{nsd2le}
  \Big[w^{-7}\partial_{w}(w^{7}\partial_{w}) + w^{-2}
  \Lambda(\lambda) \Big] H_{2}(w,\lambda) \ = \  0
\end{equation}
The solution with no $\lambda$ dependence is $H_{2}=1+Q_{2}/w^{6}$,
thus getting the following SUGRA solution
\begin{eqnarray}
  \label{nsd2me}
  ds^{2} & = &  \frac{w^{3}}{\sqrt{Q_{2}}} dx_{01}^{2} +
  \frac{\sqrt{Q_{2}}}{w^{3}} dw^{2} + \frac{\sqrt{Q_{2}}}{w} ( d \lambda^{2} +
  \cos^{2} \lambda d \Omega_{3}^{2} + \frac{1}{4} \sin^{2} \lambda d
  \Omega_{2}^{2} ) + \frac{4Q_{5}^{2}w}{\sqrt{Q_{2}} \sin^{2} \lambda}
    d x_{6}^{2} \nonumber \\
e^{2 (\phi-\phi_{\infty})} & = & H_{5} H_{2}^{1/2} = \frac{4Q_{5}^{2}
  \sqrt{Q_{2}}}{w^{5} \sin^{2} \lambda} \nonumber \\
R & \sim & \frac{w}{\sqrt{Q_{2}}} 
\end{eqnarray}
where now $R$ is the curvature on the $(01w)$ part of the metric.
The field theory is a 1+1 dimensional $\Pi_{i=1}^{N_{5}}SU(N_{2})$ gauge
theory, with an extra $U(1)$, and with $N_{5}$
hyper-multiplets, and 8 supersymmetries. It can be described by
perturbation theory when $W \gg g_{YM} \sqrt{N_{2}}$. We can express $Q_{2}, Q_{5}$
in terms of field theory variables as
\begin{equation}
  \label{q2q5}
  Q_{2} \sim g_{YM}^{2} N_{2} \ap^{4} \ \ \ \ \ Q_{5} \sim N_{5}
  \ap^{1/2} \ \ \ \ \ e^{2 \phi_{\infty}}=g_{s}^{2} \sim g_{YM}^{4} \ap^{2}
\end{equation}
Thus the geometric description will be valid when 
\begin{equation}
  \label{nsd2con}
  \Big[ \frac{N_{5}^{2} g_{YM}^{5} \sqrt{N_{2}}}{\sin^{2} \lambda} \Big] ^{1/5}
  < W < g_{YM} \sqrt{N_{2}}  
\end{equation}
The solution can be elevated to M-theory, where we find an intersection
of a localized M2 and an M5 smeared along (6,11). The solutions is
described by the same harmonic functions 
\begin{eqnarray}
  \label{nsd211r}
ds^{2} & = & H_{2}^{-2/3}H_{5}^{-1/3} dx_{01}^{2} +
  H_{2}^{1/3}H_{5}^{-1/3}dx_{2345}^{2}+ H_{2}^{-2/3}H_{5}^{2/3}
  dx_{6,10}^{2} +H_{2}^{1/3}H_{5}^{2/3} dx_{789}^{2}
\nonumber \\
  l_{p}^{2} R & \sim &  \Big[ \frac{N_{5}g_{YM}}{W \sqrt{N_{2}} \sin
    \lambda} \Big] ^{2/3} 
\end{eqnarray}
There is also the T-dual description of a D1 on an orbifold
singularity given by
\begin{eqnarray}
  \label{nsd2td}
  ds^{2} & = & \frac{w^{3}}{\sqrt{Q_{2}}} dx_{01}^{2} +
\frac{\sqrt{Q_{2}}}{w^{3}} d w^{2} + \frac{\sqrt{Q_{2}}}{w} ds_{X}^{2} \nonumber \\ 
e^{2 \phi} & = & H_{2} \sim \frac{g_{YM}^{6} N_{2}}{W^{6}}
\end{eqnarray}
The T-dual solution resembles the solution found for a D1 brane in flat space
\cite{cobi}, with the original $d \Omega_{7}$ orbifolded. 
\subsection{NS $\bot$ D1}
\label{sec:nsd1}
The last configuration that we describe in this section has
no apparent field theory dual. A D1-brane along the 6 direction intersecting a
smeared NS brane is described by the function $H_{1}(v,r)$ that obeys
\begin{equation}
  \label{nsd1lp}
  \Big[ \del_{12345} +H_{5}^{-1} \del_{789} \Big] H_{1} (v,r) \ = \ 0
\end{equation}
Using similar techniques as in the previous cases, we change variables
to $w,\lambda$ and get the following equation
\begin{equation}
  \label{nsd1lp1}
  \Big[ w^{-8} \partial_{w}(w^{8} \partial_{w}) + w^{-2} \Lambda
  (\lambda) \Big] H_{1}(w, \lambda) \ = \ 0
\end{equation}
A solution with no $\lambda$ dependence gives the following metric 
\begin{eqnarray}
  \label{nsd1met}
  ds^{2} & = & \frac{w^{7/2}}{\sqrt{Q_{1}}} dx_{01}^{2} +
  \frac{\sqrt{Q_{1}}}{w^{7/2}} dw^{2} + \sqrt{Q_{1}} ( d \lambda^{2} +
  \cos^{2} \lambda d \Omega^{2}_{4} +\frac{1}{4} \sin^{2} \lambda d
  \Omega^{2}_{2} ) + \frac{4Q_{5}^{2}w^{3/2}}{\sqrt{Q_{1}} \sin^{2}
    \lambda} dx_{6}^{2}
 \nonumber \\ e^{2 (\phi-\phi_{\infty})} & = & H_{5} H_{1} =
 \frac{4Q_{5}^{2}Q_{1}}{w^{9} \sin^{2} \lambda}
\end{eqnarray}
As in the case of a D3 $\bot$ NS, we can map this solution to its
S-dual by multiplying the metric by $H_{5}^{-1/2}$ and taking $\phi
\rightarrow -\phi$. We can also T-dualize to get a D0 smeared in the
1 direction and localized on an orbifold singularity. In \cite{cobi}
it was speculated that the solution of N D0 is dual to super-symmetric
quantum mechanics. 

\section{Semi-localized Dp branes}
\label{sec:3}
We now turn to configurations in which the Dp brane is smeared
in the world volume of the NS brane. We shall find an explicit
solution for D1 $\bot$ NS. Solutions for Dp-branes up to p=4 will follow by T duality.

In the following construction the NS brane spans (12345), and is
completely localized in (6789). The D1 (6) is localized in (789), but is
smeared in (12345). Thus the harmonic function of the D1 brane does
not depend on $x_{12345}$ and should obey $\del_{789} H_{1} =0$.
This gives $H_{1}=1+Q_{1}/r$. In the near horizon limit we
neglect the 1 in $H_{1}$. Thus $H_{5}$ satisfies a curved space Laplace
equation
\begin{equation}
  \label{nsdle}
  \Big[ \del_{6} + H_{1}^{-1} \del_{789} \Big] H_{5}(x,r) \ = \ 0
\end{equation}
By defining $u=2 \sqrt{Q_{1}r}$ we get
\begin{equation}
  \label{nsdle1}
  \Big[ \partial_{6}^{2} + u^{-3} \partial_{u}(u^{3} \partial_{u}) \Big] H_{5}(x,u) \ = \ 0
\end{equation}
Eq.(\ref{nsdle1}) can be written using $w^{2}=x_{6}^{2}+u^{2}$ and
$\tan \lambda = u/x_{6}$. 
\begin{equation}
  \label{nsdle2}
  \Big[  w^{-4} \partial_{w}(w^{4} \partial_{w}) + w^{-2} \Lambda(\lambda) \Big] H_{5}(w,\lambda) \ = \ 0
\end{equation}
We shall concentrate on solutions that have no $\lambda$ dependence,
$H_{5}=1+Q_{5}/w^{3}$. This solution can be generalized to
\begin{equation}
  \label{nsdg}
  H_{5} \ = \ 1+ \sum_{i=1}^{N_{5}} \frac{\mu_{i}}{((x-x_{i})^{2} + 4Q_{1}r)^{3/2}}
\end{equation}
The full SUGRA solution is according to eq.(\ref{hsr}) :
\begin{eqnarray}
  \label{nsdmet}
  ds^{2} & = & \frac{w \sin \lambda}{2Q_{1}} dx_{0}^{2} + \frac{2Q_{1}}{w
    \sin \lambda} d x_{12345}^{2} + \frac{Q_{5} \sin \lambda }{2 Q_{1}
    w^{2}} dx_{6}^{2} + \frac{2Q_{1}Q_{5}}{w^{4} \sin \lambda} dx_{789}^{2} \nonumber \\
e^{2 (\phi-\phi_{\infty})} & = & H_{5} H_{1} = \frac{4Q_{1}^{2}Q_{5}}{w^{5} \sin^{2} \lambda}
\end{eqnarray}
It can be seen that the procedure will be the same for all other Dp
branes up to p=4. This is equivalent to T-dualizing along (12345). The
effect of this duality is just $g_{ii} \rightarrow 1/g_{ii}$. The case
of an NS localized within a D6 brane was considered from another point
of view in \cite{sanny,oscar}. It is also related by T-duality to the
solutions shown here. Since in these solution the NS-brane is
localized, we expect the corresponding field theory to live on the
NS-brane. This field theory will have a RR background from the
Dp-branes, and 8 supersymmetries. We shall not explore such field
theories here.

\section{Summary and discussion}
\label{sec:5}

We have displayed solutions of semi-localized intersecting NS and
Dp-branes. Given the metrics of these intersections, one can learn
about the field theory duals to these configurations by calculating
Wilson loops. For example, When calculating the Wilson loop in the NS
$\bot$ D4 case, the metric factors into $AdS_{5} \times X^{5}$. Thus,
provided that we do not move on $X^{5}$, i.e. change$\lambda$, the
result will be independent of the nature of $X^{5}$. The same is also true for the rest of the configurations,
that the Wilson loop of the field theory is determined by the part in the
metric corresponding to the brane world volume coordinates, while the
number of supersymmetries and matter content are determined by the
nature of $X$. The solutions for Dp-branes intersecting NS
branes were shown to be dual to D(p-1)-branes on singular
backgrounds.

Similar solutions
for Dp-branes intersecting Dp-branes can be found using the techniques
shown here. It is, however, not clear what field theories a general
intersection describes.

\paragraph{Acknowledgments.}
I would like to thank my supervisors Prof. J. Sonnenschein and
Prof. S. Yankielowicz for helpful discussions. I would also like to
thank N. Itzhaki, D. Smith, and A. Fayyazuddin for their comments.




\begin{thebibliography}{99}
\bibitem{gaun} Jerome P. Gauntlett, ``Intersecting Branes'',
  hep-th/9705011, and references therein
\bibitem{tow} J. P. Gauntlett, G. W. Gibbons, G. Papadopoulos,
  P. K. Townsend, ``Hyper-Kahler manifolds and multiply-intersecting
  branes'', Nucl.Phys. B500 (1997) 133-162, hep-th/9702202
\bibitem{tow1}  Joaquim Gomis, David Mateos, Joan Simón, Paul
  K. Townsend, ``Brane-Intersection Dynamics from Branes in Brane
  Backgrounds'', Phys.Lett. B430 (1998) 231-236, hep-th/9803040  
\bibitem{tsey} A. A. Tseytlin, ``Composite BPS configurations of
  p-branes in 10 and 11 dimensions'', Class.Quant.Grav. 14 (1997)
  2085-2105, hep-th/9702163
\bibitem{tsey1} A. A. Tseytlin, ``No-force condition and BPS
  combinations of p-branes in 11 and 10 dimensions'', Nucl.Phys. B487
  (1997) 141-154, hep-th/9609212
\bibitem{tsey2} A. A. Tseytlin, ``Harmonic superpositions of
  M-branes'', Nucl.Phys. B475 (1996) 149-163, hep-th/9604035
\bibitem{englert}  R. Argurio, F. Englert, L. Houart, ``Intersection Rules
  for p-Branes'', Phys.Lett. B398 (1997) 61-68, hep-th/9701042
\bibitem{boonstra} Harm Jan Boonstra, Bas Peeters, Kostas Skenderis,
  ``Brane intersections, anti-de Sitter spacetimes and dual
  superconformal theories'', Nucl.Phys. B533 (1998) 127-162,
  hep-th/9803231
\bibitem{sanny}  N. Itzhaki, A. A. Tseytlin, S. Yankielowicz,
  ``Supergravity Solutions for Branes Localized Within Branes'',
  Phys.Lett. B432 (1998) 298-304, hep-th/9803103
\bibitem{hashimoto}  Akikazu Hashimoto, ``Supergravity Solutions for
  Localized Intersections of Branes'', hep-th/9812159
\bibitem{tatar}  Jose D. Edelstein, Liviu Tataru, Radu Tatar, ``Rules
  for Localized Overlappings and Intersections of p-Branes'', JHEP
  9806 (1998) 003, hep-th/9801049
\bibitem{bert}  Klaus Behrndt, Eric Bergshoeff, Bert Janssen,
  ``Intersecting D-Branes in ten and six dimensions'', Phys.Rev. D55
  (1997) 3785-3792, hep-th/9604168
\bibitem{oscar}  Oskar Pelc, Ruud Siebelink, ``The D2-D6 System and a
  Fibered AdS Geometry'', hep-th/9902045
\bibitem{witten}  Amihay Hanany, Edward Witten, ``Type IIB
  Superstrings, BPS Monopoles, And Three-Dimensional Gauge Dynamics'',
  Nucl.Phys. B492 (1997) 152-190, hep-th/9611230
\bibitem{witten1}  Edward Witten, ``Solutions Of Four-Dimensional
  Field Theories Via M Theory'', Nucl.Phys. B500 (1997) 3-42, hep-th/9703166
\bibitem{gk} A. Giveon, D. Kutasov, `` Brane Dynamics and Gauge
  Theory'',  hep-th/9802067
\bibitem{berg}  E. Bergshoeff, C. M. Hull, T. Ortin, ``Duality in the
  Type--II Superstring Effective Action'', Nucl.Phys. B451 (1995)
  547-578, hep-th/9504081
\bibitem{malda}  Juan M. Maldacena, ``The Large N Limit of Superconformal Field Theories and Supergravity
     Authors: Juan M. Maldacena'', Adv.Theor.Math.Phys. 2 (1998)
     231-252, hep-th/9711200
\bibitem{cobi} Nissan Itzhaki, Juan M. Maldacena, Jacob Sonnenschein,
  Shimon Yankielowicz, ``Supergravity and The Large N Limit of
  Theories With Sixteen Supercharges'',  Phys.Rev. D58 (1998) 046004, hep-th/9802042
\bibitem{kachru}  S. Kachru, E. Silverstein, ``4d Conformal Field
  Theories and Strings on Orbifolds'', Phys.Rev.Lett. 80 (1998)
  4855-4858, hep-th/9802183
\bibitem{vafa}  A. Lawrence, N. Nekrasov, C. Vafa, ``On Conformal
  Theories in Four Dimensions'', Nucl.Phys. B533 (1998) 199-209,
  hep-th/9803015
\bibitem{dasgupta} Keshav Dasgupta, Sunil Mukhi, ``Brane
  Constructions, Conifolds and M-Theory'',  hep-th/9811139 
\bibitem{uranga} Angel M. Uranga, ``Brane Configurations for Branes at
  Conifolds'',  hep-th/9811004
\bibitem{yang} Haisong Yang, ``Localized Intersecting Brane Solutions
  of D=11 Supergravity'', hep-th/9902128

\end{thebibliography}
\end{document}